\documentclass[showpacs,prd,twocolumn,aps]{revtex4}

\usepackage{bm,graphicx,textcomp,amssymb,amsmath,dcolumn,hyperref,color}

\newcommand{\LCperp}{{\scriptscriptstyle \perp}}

\newcommand{\ket}[1]{\left|#1\right>}

\newcommand{\nn}{\nonumber\\}

\newcommand{\f}[1]{\mbox{\boldmath$#1$}}
\newcommand{\fk}[1]{\mbox{\boldmath$\scriptstyle#1$}}

\newcommand{\bea}{\begin{eqnarray}}
\newcommand{\ea}{\end{eqnarray}}
\newcommand{\eea}{\end{eqnarray}}

\newcommand{\be}{\begin{equation}}
\newcommand{\ee}{\end{equation}}

\begin{document}

\title{Doubly assisted Sauter-Schwinger effect}

\author{Greger Torgrimsson}
\author{Johannes Oertel}
\author{Ralf Sch\"utzhold}
\affiliation{Fakult\"at f\"ur Physik, 
Universit\"at Duisburg-Essen, Lotharstra{\ss}e 1, Duisburg 47048, Germany}

\date{\today}

\begin{abstract}
We study electron-positron pair creation by a strong and slowly 
varying electric field, assisted by a weaker and more rapidly changing field 
(e.g., in the keV regime) plus an additional high-energy (say MeV) photon. 
It turns out that this combination can yield a pair creation probability 
which is exponentially larger than in the case where one (or more) of the 
three ingredients is missing. 
Apart from a deeper understanding of these non-perturbative phenomena,
this double enhancement may pave the way for an experimental verification
of this fundamental prediction. 
\end{abstract}


\maketitle

\section{Introduction}

One of the fundamental predictions of quantum field theory which has so far 
resisted a direct experimental verification is the Sauter-Schwinger effect,
i.e., the creation of electron-positron pairs out of the quantum vacuum due
to a strong electric field~\cite{Sauter:1931zz,Heisenberg:1935qt,Schwinger:1951nm}.
In order to observe this striking phenomenon, the electric field should not be 
too far below the Schwinger critical field $E_S=m^2c^3/(\hbar q)$ which is 
determined by the electron charge $q$ and mass $m$ and thus quite large 
$E_S\approx 1.3\times 10^{18}\text{V/m}$.
For a constant electric field $E$, the leading-order probability per unit 
time and volume is given by ($\hbar=c=1$) 
\bea
\label{probability}
\frac{P_{e^+e^-}}{\Delta t\Delta V}
=
\frac{q^2E^2}{4\pi^3}\,\exp\left\{-\frac{\pi m^2}{qE}\right\} 
\,.
\ea
For example, an optical laser has a typical focus size of 
$\Delta t\Delta V\sim(\mu\text{m})^4$, and thus one would need a field 
strength of $E\approx8\times10^{16}\text{V/m}$, 
i.e., six percent of the critical field $E_S$,  
in order to have a probability of order one.
Translated into an intensity, this corresponds to 
$I=\epsilon_0E^2\approx2\times10^{27}\text{W}/\text{cm}^2$, 
see also~\cite{Bulanov:2004de}.
Even though this magnitude is still beyond present day technology, 
there are several experimental initiatives which intend to reach ultra-high 
field strengths coming excitingly close to this value. 

This motivates the search for enhancement mechanisms which might reduce 
the required field strength.
As one possible scenario, it has been found \cite{Schutzhold:2008pz} 
that the superposition of a 
strong and constant electric field $E$ with a weaker time-dependent field, 
such as 
\bea
\label{field}
\f{E}(t)=E\left[1+\varepsilon f(t)\right]\f{e}_z
\,,
\ea
where $f(t)$ could be a Sauter pulse $f(t)=1/\cosh^2(\omega t)$, for example,
can induce an exponentially enhanced pair-creation probability 
({\em dynamically assisted Sauter-Schwinger effect}, 
see e.g.~\cite{Schutzhold:2008pz,Orthaber:2011cm,Otto:2015gla}).
Here, we consider even functions $\f{E}(-t)=\f{E}(t)$ for simplicity. 
Since the relevant parameter here is the combined Keldysh parameter 
$\gamma=m\omega/(qE)$, this enhancement mechanism can be operative already
at frequency scales $\omega$ well below the electron mass scale $m$.

As another scenario, the combined impact of a strong and constant 
electric field $E$ plus an additional photon with wave-number $\f{K}$ 
and frequency $\Omega=|\f{K}|$ 
can also exponentially enhance the pair-creation probability 
\cite{Dunne:2009gi}. 
In contrast to the first scenario, this enhancement mechanism is only effective
for photon frequencies $\Omega$ of the order of the electron mass, 
i.e., in the MeV regime.

In the following, we consider a combination of these two scenarios, i.e.,
a strong and constant electric field $E$ plus a weaker time-dependent field, 
as in \eqref{field}, plus an additional high-energy photon. 
For example, one could envisage that $E$ represents the focus of an optical 
laser (which can be approximated by a constant field), while the weaker field
$\varepsilon E f(t)$ is generated by a focused XFEL beam 
(in the multi-keV regime).
Naively, one might expect that the interaction of this field \eqref{field}
with the high-energy photon can be treated within the locally constant field 
approximation because the photon energy (of order MeV) is much larger than the 
XFEL frequency $\omega$.
This would suggest that the weaker field has negligible effect. 
However, as we will see below, this simplified view can be misleading and 
the combined impact of all three ingredients can generate further exponential
enhancement of the pair-creation probability. 
Similar ideas to enhance the pair production probability 
with ``three ingredients'' can be found in 
\cite{DiPiazza:2009py,Nousch:2015pja,Otto:2016fdo}, 
see also~\cite{Bulanov:2010ei} and 
\footnote{Note that the doubly assisted Sauter-Schwinger effect is 
qualitatively and kinematically quite different from other phenomena such as 
the assisted Breit-Wheeler processes as studied in 
\cite{Nousch:2015pja,Otto:2016fdo}, for example. 
The interaction of a high-energy photon with a field of the form 
\eqref{field} or \eqref{standing} is very different from the interaction 
of a high-energy photon with a counter-propagating plane wave, even if 
this wave is decomposed of two or more 
frequencies~\cite{Nousch:2015pja,Otto:2016fdo}. 
For example, electrons and positrons under the influence of the plane wave 
alone can be described by Volkov solutions and there is no pair 
creation (no Sauter-Schwinger effect) without the high-energy photon.
Consistently, one cannot associate a Lorentz invariant measure to the 
field strength or the frequency of a plane wave and thus the threshold 
behavior found for the (doubly) assisted Sauter-Schwinger effect cannot 
occur. 
}. 

\section{WKB approach}

Because the electric field \eqref{field} consists of a very large number of 
coherent photons, we treat it as a classical (external) background field 
$A_\mu^{\rm class}(t)$ in temporal gauge. 
In analogy to the Furry picture, this allows us to treat this field 
\eqref{field} non-perturbatively. 
Thus, the total field operator is given by  
$\hat A_\mu(t,\f{r})=A_\mu^{\rm class}(t)+\hat A_\mu^{\rm quant}(t,\f{r})$
where $\hat A_\mu^{\rm quant}(t,\f{r})$ contains the additional high-energy 
photon.
Then, the amplitude for the  transition from an initial state with one photon 
$\ket{{\rm in}}=\hat a^\dagger_{\fk{K}}\ket{0_{\rm in}}$ 
to a final state containing an electron-positron pair can be estimated via 
time-dependent perturbation theory with the interaction Hamiltonian 
\bea
\label{H-int}
\hat H_{\rm int}
=
q\int d^3r\,\hat{\bar\psi}\,\gamma^\mu\hat A_\mu^{\rm quant}\,\hat\psi
\,,
\ea
where the fermionic field operator $\hat\psi$ satisfies the Dirac equation 
($\hat H_0$-dynamics) in the background $A_\mu^{\rm class}(t)$ 
\bea
\label{Dirac}
\left(i\gamma^\mu\partial_\mu-q\gamma^\mu A_\mu^{\rm class}-m\right)\hat\psi
=
0
\,.
\ea
Compare \cite{Wollert:2015oea}, where similar methods were used to study 
photon-stimulated pair production in constant fields, 
see also~\cite{Fedotov:2013uja,DiPiazza:2015xva}.
Since $A_\mu^{\rm class}(t)$ only depends on time, we may describe the position 
dependence of $\hat\psi$ by a spatial Fourier decomposition. 
The remaining temporal dependence is then treated via the WKB approximation 
\bea
\label{Fourier}
\hat\psi(t,\f{r})
&=&
\sum_s 
\int d^3k\left(
e^{+i\fk{k}\cdot\fk{r}}
u_{\fk{k},s}(t)
e^{-iS^+_{\fk{k}}(t)}
\hat c_{\fk{k},s}
+
\right.
\nn
&&
\left.
+
e^{-i\fk{k}\cdot\fk{r}}
v_{\fk{k},s}(t)
e^{+iS^-_{\fk{k}}(t)}
\hat d_{\fk{k},s}^\dagger
\right) 
\,,
\ea
where $\hat d_{\fk{k},s}^\dagger$ and $\hat c_{\fk{k},s}$ are the (initial) 
creation and annihilation operators 
$\hat c_{\fk{k},s}\ket{0_{\rm in}}=\hat d_{\fk{k},s}\ket{0_{\rm in}}=0$
for positrons and electrons with
wave-number $\f{k}$ and spin $s$, respectively. 
The associated spinor components $u_{\fk{k},s}(t)$ and $v_{\fk{k},s}(t)$ are
slowly varying with time, while the remaining rapid temporal oscillation is 
given by the exponential of the phase (eikonal or Hamilton-Jacobi) function 
\bea
\label{eikonal}
\dot S^\pm_{\fk{k}}(t)=\sqrt{m^2+\left[\f{k}\pm q\f{A}^{\rm class}(t)\right]^2}
\,.
\ea
Strictly speaking, there is a mixing of positive $e^{-iS^+_{\fk{k}}(t)}$ 
and negative  $e^{+iS^+_{\fk{k}}(t)}$ frequencies in front of 
$\hat c_{\fk{k},s}$ (and analogously for $\hat d_{\fk{k},s}^\dagger$) 
which reflect pair creation from the field \eqref{field} alone,
i.e., without the additional high-energy photon. 
These terms are neglected here, but we have checked that this approximation 
does not affect our main results.  

Now we may estimate the amplitude for creating an electron-positron pair with
wave-numbers $\f{k}^\pm$ by inserting \eqref{Fourier} into \eqref{H-int}.
The spatial integral in \eqref{H-int} yields $\delta^3(\f{k}^++\f{k}^--\f{K})$.
In the following, we focus on the symmetric case $\f{k}^+=\f{k}^-=\f{K}/2$ 
(c.f.~\cite{Fedotov:2013uja,Wollert:2015oea})
since it gives the largest amplitude -- it corresponds to the saddle point 
of the remaining momentum integration from \eqref{Fourier}. 
After these steps, the amplitude $\mathfrak A$ for pair creation 
is determined by a time integral 
\bea
\label{amplitude}
{\mathfrak A}
=
\int dt\,M(t)
\exp\left\{-i\Omega t+iS^+_{\fk{k}^+}(t)+iS^-_{\fk{k}^-}(t)\right\}
\,,
\ea
where $M(t)$ is a slowly varying prefactor containing the spinors
$u_{\fk{k}^+,s}^\dagger(t)$ and $v_{\fk{k}^-,s}(t)$ etc. 
The remaining exponential is rapidly oscillating (along the real time axis)
which motivates estimating this time integral by the saddle point method. 

\section{Saddle-point approximation}

The saddle points $t_*$ in the complex $t$-plane of the 
integral~\eqref{amplitude} are given by the implicit equation  
\bea
\label{saddle}
\Omega=\dot S^+_{\fk{k}^+}(t_*)+\dot S^-_{\fk{k}^-}(t_*)
\,.
\ea
Interestingly, after inserting \eqref{eikonal}, the saddle points $t_*$ 
do not depend on the frequency $\Omega$, but just on the direction of the 
photon momentum $\f{K}$ relative to $\f{E}$ field.
In the following, we consider the perpendicular case $\f{K}\perp\f{E}$ 
because it yields the largest pair creation probability, 
see also~\cite{Dunne:2009gi}.  
The dependence on the angle between $\f{K}$ and $\f{E}$ will be discussed 
elsewhere \cite{Doubly2}.
In this perpendicular case, we obtain the same values for $t_*$ as already
discussed in \cite{Linder:2015vta} for the case without the additional 
high-energy photon, i.e., just the field \eqref{field}.

\subsection{Sauter pulse}

Following \cite{Schutzhold:2008pz,Linder:2015vta}, we first discuss the case 
where the background field $A_\mu^{\rm class}(t)$ has poles at imaginary times 
$t_*=\pm i\tau_*$, e.g., where the function $f(t)$ in \eqref{field} is 
given by a Sauter pulse 
$f(t)=1/\cosh^2(\omega t)$ or a Lorentzian profile $f(t)=1/(1+\omega^2t^2)$.
Introducing the combined effective Keldysh parameter as
\bea
\label{Keldysh}
\gamma_*=\frac{m}{qE\tau_*}
\,,
\ea
we obtain the same threshold behavior as already discussed in 
\cite{Schutzhold:2008pz,Linder:2015vta}.
Below threshold $\gamma_*<1$, the ordinary saddle point $t_*=im/(qE)$ 
dominates and the weak field $\propto\varepsilon$ has negligible impact
for small $\varepsilon$.
Above threshold $\gamma_*>1$, however, the additional saddle points 
introduced by the weak field $t_*\approx\pm i\tau_*$ dominate. 
In this case, we may estimate the pair creation probability via the 
saddle point method and obtain 
\bea
\label{doubly}
P_{e^+e^-}
\sim
\exp\left\{-2\Im\left[
S^+_{\fk{k}^+}(t_*)+S^-_{\fk{k}^-}(t_*)-\Omega t_*
\right]\right\}
=
\nn
\exp\left\{
-\frac{2m_\LCperp^2}{qE}
\left[
\frac{\sqrt{\gamma_\LCperp^2-1}}{\gamma_\LCperp^2}
+\arcsin\left[\frac{1}{\gamma_\LCperp}\right]
-\frac{\Omega}{m_\LCperp\gamma_\LCperp}
\right]\right\}
,
\ea
where $m^2_\perp=m^2+\f{k}_\pm^2=m^2+(\Omega/2)^2$ denotes the effective 
transversal mass and $\gamma_\LCperp=\gamma_*m_\LCperp/m$
(remember that we are considering the perpendicular case $\f{K}\perp\f{E}$ here). 
Comparing the above result~\eqref{doubly} with Eq.~(12) in 
\cite{Schutzhold:2008pz} for dynamically assisted pair production without 
the high-energy photon, we find that the first two terms in the second line 
of~\eqref{doubly} can be obtained from Eq.~(12) in \cite{Schutzhold:2008pz} 
by replacing $m$ and $\gamma_*$ with $m_\LCperp$ and $\gamma_\LCperp$.
Since this essentially corresponds to heavier particles 
(c.f.~\cite{Wollert:2015oea}), this change alone would lead to a 
reduction of the exponential (i.e., pair-creation probability $P_{e^+e^-}$). 
However, because of the last term $\propto\Omega$ in~\eqref{doubly}, 
which stems from the $e^{-i\Omega t}$ term in~\eqref{amplitude}, 
the total exponential is larger in our doubly assisted case.

Let us discuss some limiting cases of this result \eqref{doubly}.
For vanishing photon frequency $\Omega\to0$, we recover Eq.~(12) in 
\cite{Schutzhold:2008pz} for dynamically assisted pair production by the 
electric field \eqref{field} alone.
As the Keldysh parameter approaches the threshold, $\gamma_*\downarrow1$, 
the effect of the weak field becomes negligible and we recover Eq.~(5) in 
\cite{Dunne:2009gi}. 
Thus, our doubly assisted result \eqref{doubly} includes the two ``singly'' 
assisted cases mentioned in the introduction in two different limits. 
However, our result \eqref{doubly} is in general larger than both these limits.

For very high photon frequencies $\Omega$, one might expect the locally 
constant field result $P_{e^+e^-}\sim e^{-8/(3\chi)}$ with 
$\chi=qE\Omega/m^3$,
see, e.g., Eq.~(36) in~\cite{Nikishov-Ritus}. 
Instead we find
\be
\label{high-photon-frequency}
\lim\limits_{\Omega\to\infty}\eqref{doubly}
=
\exp
\left\{
-\frac{8}{3\chi}\frac{3\gamma_*^2-1}{2\gamma_*^3}
\right\}
\geq e^{-\frac{8}{3\chi}} 
\;,
\ee
which (recall $\gamma_*>1$) is larger than the locally constant field result. 
This failure of the locally constant field approximation suggests the 
following picture: 
Even though the transversal oscillations of the photon $\Omega=|\f{K}|$ and 
the electron-positron pair $\f{k}_\pm=\f{K}/2$ are too fast to resolve the 
temporal variations in $E(t)$, the longitudinal behavior of the 
electron-positron wave-functions is much slower and thus {\em does} 
feel the time-dependence of $E(t)$. 

Finally, if the frequency scale $\omega$ of the weak field is very high, 
we find
\be
\label{high-weak-frequency}
\lim\limits_{\gamma_*\to\infty}\eqref{doubly}
=
e^{-2\tau_*\left(2m_\perp-\Omega\right)} 
\;,
\ee
which for $\Omega\to0$ agrees with Eq.~(30) in \cite{Popov72} 
(see also~\cite{Dunne:1999uy}). 
Note that this result \eqref{high-weak-frequency} is independent of the 
electric field strength $E$ and also gives the exponent for the case 
without the strong field.  
As one would expect, this exponent \eqref{high-weak-frequency} can be derived 
via perturbation theory \cite{Doubly2}. 
However, since we are mostly interested in pair creation in the non-perturbative 
regime, we want to avoid very large values of $\gamma_*$.

\subsection{Sinusoidal profile}

As another important example, let us consider an oscillating profile 
$f(t)=\cos(\omega t)$.
As the associated vector potential 
$A_\mu^{\rm class}(t)\propto[t+\varepsilon\sin(\omega t)/\omega]$ is analytic in 
the entire complex plane, the situation is qualitatively different from the 
previous subsection. 
The saddle points $t_*$ are given by the transcendental equation 
\cite{Linder:2015vta}
\bea
\label{transcendental}
\omega t_*+\varepsilon\sin(\omega t_*)=\pm i\gamma
\,,
\ea
with $\gamma=m\omega/(qE)$. 
In contrast to the previous subsection, the threshold value 
$\gamma^{\rm crit}$ now depends on the strength of the weak field 
and scales as $\gamma^{\rm crit}\sim|\ln\varepsilon|$ \cite{Linder:2015vta}.
Again, below threshold $\gamma<\gamma^{\rm crit}$, the weak field 
$\propto\varepsilon$ has negligible impact for small $\varepsilon$.
Above threshold  $\gamma>\gamma^{\rm crit}$, however, the saddle points 
in \eqref{transcendental} move towards the real time-axis and thus the 
pair creation probability is amplified exponentially.
Unfortunately, we do not obtain a simple explicit analytic solution as in 
\eqref{doubly}, but we have an implicit expression by inserting the solution 
$t_*$ of \eqref{transcendental} into the saddle-point estimate 
$P_{e^+e^-}\sim\exp\left\{-2\Im\left[
S^+_{\fk{k}^+}(t_*)+S^-_{\fk{k}^-}(t_*)-\Omega t_*
\right]\right\}$, which can easily be evaluated 
numerically, see Fig.~\ref{ES-strong-cos-photon1-fig}.

\section{World-line instantons}

Instead of the WKB approach sketched above, we can also derive the 
pair-creation probability via the world-line instanton formalism 
\cite{Affleck:1981bma,Dunne:2005sx,Dunne:2006st,Dunne:2006ur}, 
which has proven useful for studying non-perturbative pair production 
in many cases \cite{Affleck:1981bma,Dunne:2005sx,Dunne:2006st,Dunne:2006ur,
Schutzhold:2008pz,CPL,Schneider:2014mla,Linder:2015vta,Ilderton:2015qda,
Dumlu:2015paa,Copinger:2016llk}, see also~\cite{Ilderton:2014mla}\footnote{In~\cite{Ilderton:2014mla} the worldline formalism was used to obtain the exact pair production 
probability for a large class of fields.} 
(It is also interesting to compare these worldline instantons with the 
semiclassical tunneling trajectories in~\cite{Wollert:2015oea}.) 
This formalism was used in \cite{Monin:2010qj,Satunin:2013an} to obtain the 
probability of photon stimulated pair production in constant fields. 

To this end, let us consider the transition amplitude $\Gamma$ for an initial 
photon with wave vector $K_\mu$ and polarization $\epsilon_\mu$ 
transforming into a final photon with $K_\mu'$ and $\epsilon_\mu'$.
We consider for simplicity scalar QED here, which gives the same exponent 
as spinor QED.
In the world-line formulation, this amplitude can be expressed 
as \cite{Schubert:2000yt,Schubert:2001he,Gies:2011he} 
\bea
\label{scalar-Gamma}
\Gamma=q^2
\int\limits_0^\infty\frac{dT}{T}
\oint\mathcal{D}x^\mu
\int\limits_0^1d\tau_1
\int\limits_0^1d\tau_2
\,
\epsilon_\mu\dot{x}^\mu(\tau_2)\epsilon'_\nu\dot{x}^\nu(\tau_1)e^{-i{\cal A}} 
\,,
\ea
where $\mathcal{D}x^\mu$ denotes the path integral over all periodic 
$x^\mu(1)=x^\mu(0)$ trajectories $x^\mu(\tau)$ parametrized by $\tau$ and 
$\dot{x}^\mu=dx^\mu/d\tau$. 
The action $\cal A$ is given by
\bea
\label{scalar-action}
{\cal A}
=
\frac{Tm^2}{2}+\int\limits_0^1d\tau
\left(
\frac{\dot{x}_\mu\dot{x}^\mu}{2T}+A_\mu^{\rm class}\dot{x}^\mu+J_\mu x^\mu
\right)
\,,
\ea
with the ``current'' 
$J_\mu(\tau)=K_\mu\delta(\tau-\tau_2)-K'_\mu\delta(\tau-\tau_1)$. 

Roughly speaking, $\Gamma$ corresponds to a polarization and wave-vector 
dependent index of refraction. 
It can be used to study vacuum birefringence and related phenomena~\footnote{
The worldline representation~\eqref{scalar-Gamma} was evaluated 
numerically with Monte Carlo techniques in~\cite{Gies:2011he} to study the 
phase velocity shift in inhomogeneous magnetic fields, and exactly 
in~\cite{Ilderton:2016qpj} for vacuum birefringence in plane wave backgrounds.}. 
Here, we are interested in the imaginary part of $\Gamma$ in forward direction, 
i.e., for $K_\mu=K_\mu'=(\Omega,\f{K})$ and $\epsilon_\mu=\epsilon_\mu'$, 
which, in analogy to 
the optical theorem, describes photon absorption into an electron-positron 
pair~\cite{Dunne:2009gi,Monin:2010qj,Satunin:2013an,Dinu:2013gaa,Meuren:2014uia}.

In order to compute $\Im\Gamma$, we again employ the saddle-point method
\cite{Monin:2010qj,Satunin:2013an}.
The saddle point of the first integral in \eqref{scalar-Gamma} over $T$ 
(usually called proper time) is given by 
$T^2_*=\int\limits_0^1d\tau\,\dot{x}_\mu\dot{x}^\mu/m^2$.
The saddle points of the path integral over $\mathcal{D}x^\mu$ are solutions
of the Lorentz type equation $\ddot{x}^\mu=T(F^{\mu\nu}\dot{x}_\nu+J^\mu)$
respecting the periodicity constraint $x^\mu(1)=x^\mu(0)$ and are called
world-line instantons.
For the remaining $\tau_{1,2}$-integrations, we may exploit the translational 
invariance in $\tau$~\cite{Schubert:2000yt,Satunin:2013an}, which implies that 
the integrand only depends on $\tau_-=\tau_2-\tau_1$.
As expected from symmetry considerations, the saddle point for $\tau_-$
is at $\tau_-=1/2$.  

The calculation is similar to that in~\cite{Dunne:2006st} for pair production 
without a stimulating photon. 
Here the instanton oscillates between two turning points given by the saddle 
points $\pm t_*$ above, and by changing variable from $\tau$ to $t(\tau)$ 
(c.f.~\cite{Dunne:2006st,Ilderton:2015qda}), we can calculate the exponent 
(for these purely time-dependent fields) without using the explicit instanton 
solution.
Altogether, we find $\Im\Gamma\sim\exp\{-{\cal A}_{\rm inst}\}$ where 
${\cal A}_{\rm inst}$ is the action \eqref{scalar-action} evaluated at the 
saddle point, i.e., at the world-line instanton trajectory. 
Comparison with the previous section reveals that we obtain the same 
exponent $P_{e^+e^-}\sim\exp\left\{-2\Im\left[
S^+_{\fk{k}^+}(t_*)+S^-_{\fk{k}^-}(t_*)-\Omega t_*
\right]\right\}$ as in the WKB approach.

However, apart from recovering our earlier results obtained via the WKB 
approach, the world-line instanton allows us to go beyond the scenario 
considered above.
Similar to~\cite{Linder:2015vta}, let us consider a standing wave profile 
\bea
\label{standing}
\f{E}(t,x)=E\left[1+\varepsilon\cos(\omega t)\cos(\omega x)\right]\f{e}_z
\,,
\ea
plus the associated magnetic field $\f{B}(t,x)$. 
If the additional high-energy photon is incident perpendicularly 
$\f{K}=K\f{e}_y$, we find instanton solutions which stay confined to 
a plane of constant $x$ if that value of $x$ corresponds to a maximum 
(or minimum) of the electric field, such as $x=0$.
Thus, these instanton trajectories are the same as in the case of a
purely time-dependent field \eqref{field}, i.e., they do not ``feel'' 
the transversal spatial dependence $\cos(\omega x)$ at all.
As a consequence, we find the same instanton action ${\cal A}_{\rm inst}$
and thus the same pair-creation exponent 
$\Im\Gamma\sim\exp\{-{\cal A}_{\rm inst}\}$
as in the purely time-dependent case. 

\section{Conclusions}

\begin{figure}
\includegraphics[width=.45\textwidth]{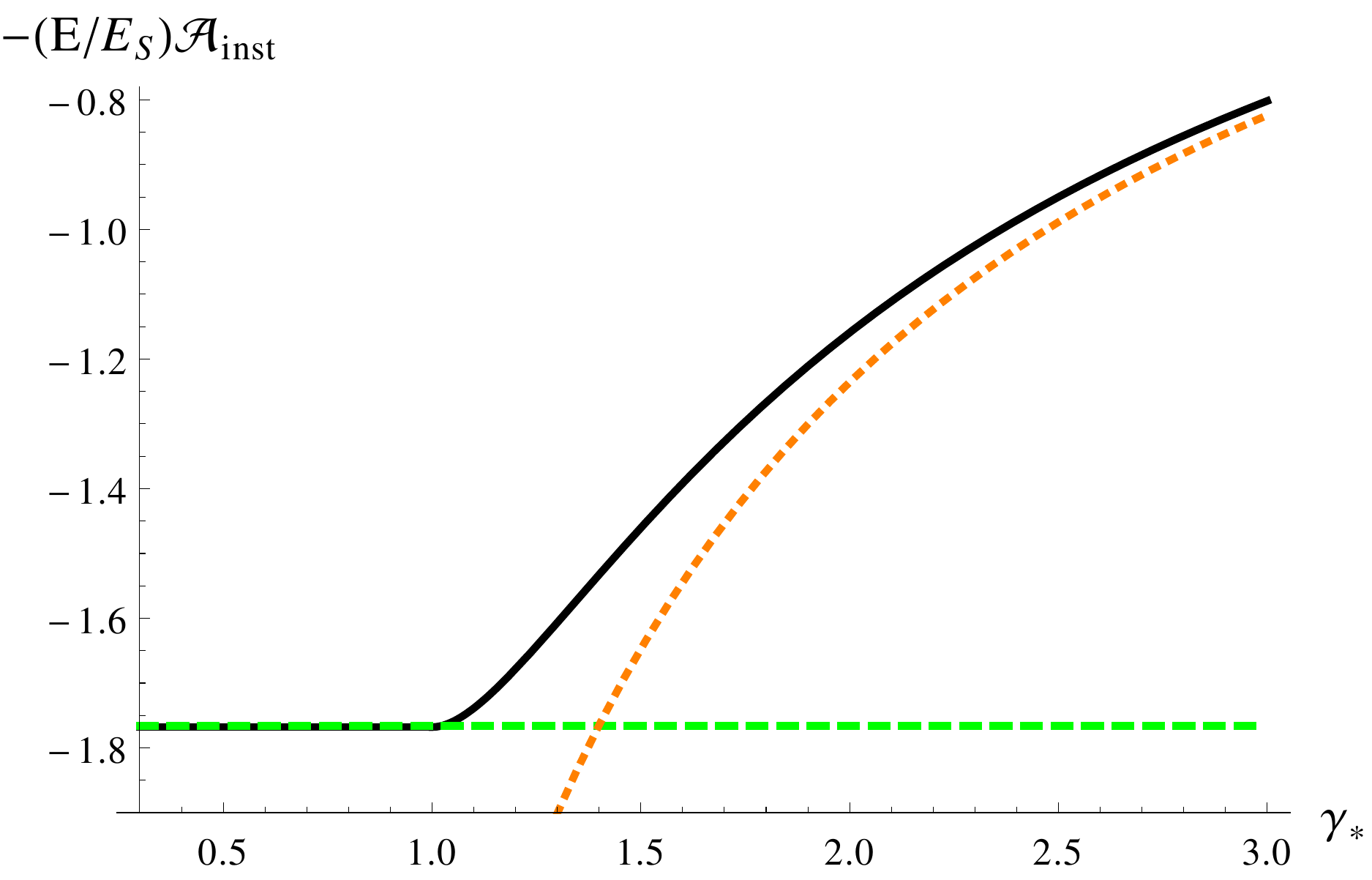}
\caption{(Color online) The solid black curve shows the instanton action 
${\cal A}_{\rm inst}$, i.e., the exponent in the pair-creation probability 
$P_{e^+e^-}\sim e^{-{\cal A}_{\rm inst}}$, in units of $E_S/E$ as a function 
of the effective Keldysh parameter $\gamma_*$ for $\Omega=m$ and a Sauter 
pulse, see Eq.~\eqref{doubly}. 
%
%
The horizontal dashed green line corresponds to photon-stimulated pair 
production in a constant electric field (i.e., the limit $\gamma_*\to 0$)
while the dotted orange curve corresponds to pair production by the photon 
and the weak field alone, which is practically identical to the field 
strength independent exponent~\eqref{high-weak-frequency}.
For intermediate values of $\gamma_*$ above threshold $\gamma_*>1$, the 
black curve (doubly assisted Sauter-Schwinger effect) lies significantly 
above the other curves where one ingredient is missing. (Note that with only the
strong field $-(E/E_S){\cal A}_{\rm inst}=-\pi$, as in~\eqref{probability}.)
%
}
\label{ES-strong-sech2-photon1-fig}
\end{figure}

\begin{figure}
\includegraphics[width=.45\textwidth]{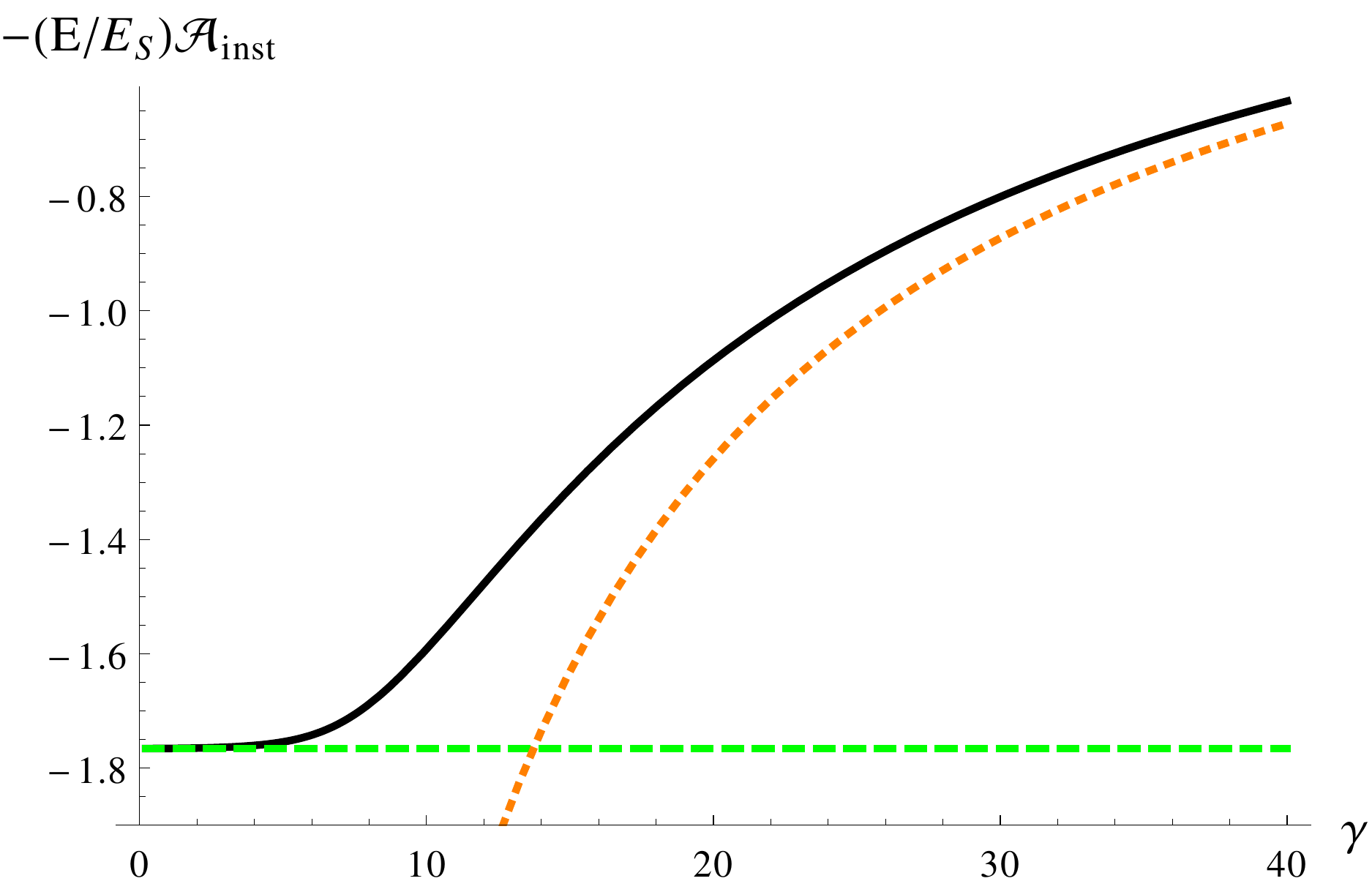}
\caption{(Color online) Analogous to Fig.~\ref{ES-strong-sech2-photon1-fig}, 
but for a sinusoidal weak field instead of a Sauter pulse 
(with the same value for $\Omega=m$). 
Compared to the Sauter pulse, the threshold is smoother and depends on the 
strength $\varepsilon=10^{-3}$ of the weak field: 
we see that the weak field starts to become important around 
$\gamma\sim|\ln\varepsilon|\approx7$, as expected.}
\label{ES-strong-cos-photon1-fig}
\end{figure}

Starting with the Sauter-Schwinger effect in a strong and constant field $E$, 
we find that the pair creation probability $P_{e^+e^-}$
can be enhanced exponentially by the combined impact of a weak time-dependent 
field plus a high-energy photon in addition to the strong and constant field. 
This {\em doubly assisted Sauter-Schwinger effect} 
is accompanied by
a failure of the locally constant field approximation. 
Similar to previous considerations, we studied weak fields in the shape of a 
Sauter pulse (or other profiles 
with poles at imaginary times, such as a Lorentzian) 
as well as oscillating behavior and found qualitative differences,
see Figs.~\ref{ES-strong-sech2-photon1-fig} and 
\ref{ES-strong-cos-photon1-fig}. 
%
%

Comparing two alternative approaches, the WKB method and the world-line 
instanton technique, we confirm that they yield the same exponent for the 
pair creation probability $P_{e^+e^-}$.
The latter approach also allows us to generalize our results to standing 
waves (i.e., vacuum solutions of the Maxwell equations) and we find that they 
yield the same exponent as the purely time-dependent case, 
see also~\cite{Linder:2015vta}.
(The prefactor in front of the exponential, which is not considered here,
will probably be different.) 
Note that we are still in the nonperturbative regime (with respect to the 
electric field), in contrast to perturbative 
(multi-photon) pair creation, as observed in the famous SLAC experiment
\cite{Bamber:1999zt}. 



\begin{thebibliography}{99}

\bibitem{Sauter:1931zz}
  F.~Sauter,
  Z.\ Phys.\  {\bf 69} (1931) 742.

\bibitem{Heisenberg:1935qt}
  W.~Heisenberg and H.~Euler,
  Z.\ Phys.\  {\bf 98} (1936) 714

\bibitem{Schwinger:1951nm}
  J.~S.~Schwinger,
  Phys.\ Rev.\  {\bf 82} (1951) 664.

\bibitem{Bulanov:2004de}
  S.~S.~Bulanov, N.~B.~Narozhny, V.~D.~Mur and V.~S.~Popov,
  Phys.\ Lett.\ A {\bf 330} (2004) 1
  [hep-ph/0403163].

\bibitem{Schutzhold:2008pz}
  R.~Sch\"utzhold, H.~Gies and G.~Dunne,
  Phys.\ Rev.\ Lett.\  {\bf 101} (2008) 130404
  [arXiv:0807.0754 [hep-th]].

\bibitem{Orthaber:2011cm}
  M.~Orthaber, F.~Hebenstreit and R.~Alkofer,
  Phys.\ Lett.\ B {\bf 698} (2011) 80
  [arXiv:1102.2182 [hep-ph]].

\bibitem{Otto:2015gla}
  A.~Otto, D.~Seipt, D.~Blaschke, S.~A.~Smolyansky and B.~K\"ampfer,
  Phys.\ Rev.\ D {\bf 91} (2015) no.10,  105018
  [arXiv:1503.08675 [hep-ph]].

\bibitem{Dunne:2009gi}
  G.~V.~Dunne, H.~Gies and R.~Sch\"utzhold,
  Phys.\ Rev.\ D {\bf 80} (2009) 111301
  [arXiv:0908.0948 [hep-ph]].

\bibitem{Nousch:2015pja}
  T.~Nousch, D.~Seipt, B.~K\"ampfer and A.~I.~Titov,
  Phys.\ Lett.\ B {\bf 755} (2016) 162
  [arXiv:1509.01983 [physics.plasm-ph]].

\bibitem{Otto:2016fdo}
  A.~Otto, T.~Nousch, D.~Seipt, B.~K\"ampfer, D.~Blaschke, A.~D.~Panferov, S.~A.~Smolyansky and A.~I.~Titov,
  J.\ Plasma Phys.\  {\bf 82} (2016) 655820301
  [arXiv:1604.00196 [hep-ph]].
  
\bibitem{DiPiazza:2009py}
  A.~Di Piazza, E.~Lotstedt, A.~I.~Milstein and C.~H.~Keitel,
  Phys.\ Rev.\ Lett.\  {\bf 103} (2009) 170403
  [arXiv:0906.0726 [hep-ph]].

\bibitem{Bulanov:2010ei}
  S.~S.~Bulanov, V.~D.~Mur, N.~B.~Narozhny, J.~Nees and V.~S.~Popov,
  Phys.\ Rev.\ Lett.\  {\bf 104} (2010) 220404
  [arXiv:1003.2623 [hep-ph]].

\bibitem{Wollert:2015oea}
  A.~W\"ollert, M.~Klaiber, H.~Bauke and C.~H.~Keitel,
  Phys.\ Rev.\ D {\bf 91} (2015) no.6,  065022.

\bibitem{Fedotov:2013uja}
  A.~M.~Fedotov and A.~A.~Mironov,
  Phys.\ Rev.\ A {\bf 88} (2013) no.6,  062110
  [arXiv:1310.7258 [hep-ph]].

\bibitem{DiPiazza:2015xva}
  A.~Di Piazza,
  Phys.\ Rev.\ A {\bf 91} (2015) no.4,  042118
  [arXiv:1501.06475 [hep-ph]].

\bibitem{Doubly2}
G.~Torgrimsson {\em et al}, in preparation. 
  
\bibitem{Linder:2015vta}
  M.~F.~Linder, C.~Schneider, J.~Sicking, N.~Szpak and R.~Sch\"utzhold,
  Phys.\ Rev.\ D {\bf 92} (2015) no.8,  085009
  [arXiv:1505.05685 [hep-th]].
  
\bibitem{Nikishov-Ritus}
A.~I.~Nikishov, V.~I.~Ritus, Sov.\ Phys.\ JETP {\bf 19} 529 (1964)  

\bibitem{Popov72}
V.~S.~Popov,
Sov.\ Phys.\ JETP {\bf 35}, 659 (1972)

\bibitem{Dunne:1999uy}
  G.~V.~Dunne and T.~M.~Hall,
  Phys.\ Rev.\ D {\bf 60} (1999) 065002
  [hep-th/9902064].



\bibitem{Dunne:2005sx}
  G.~V.~Dunne and C.~Schubert,
  Phys.\ Rev.\ D {\bf 72} (2005) 105004
  [hep-th/0507174].


\bibitem{Affleck:1981bma}
  I.~K.~Affleck, O.~Alvarez and N.~S.~Manton,
  Nucl.\ Phys.\ B {\bf 197} (1982) 509.

\bibitem{Dunne:2006st}
  G.~V.~Dunne, Q.~h.~Wang, H.~Gies and C.~Schubert,
  Phys.\ Rev.\ D {\bf 73} (2006) 065028
  [hep-th/0602176].

\bibitem{Dunne:2006ur}
  G.~V.~Dunne and Q.~h.~Wang,
  Phys.\ Rev.\ D {\bf 74} (2006) 065015
  [hep-th/0608020].

\bibitem{CPL}
X.~Bai-Song, M.~Melike, D.~Sayipjamal,
Chinese Physics Letters {\bf 29} (2012) 021102

\bibitem{Ilderton:2015qda}
  A.~Ilderton, G.~Torgrimsson and J.~W\aa rdh,
  Phys.\ Rev.\ D {\bf 92} (2015) no.6,  065001
  [arXiv:1506.09186 [hep-th]].
  
\bibitem{Copinger:2016llk}
  P.~Copinger and K.~Fukushima,
  arXiv:1605.05957 [hep-th].

\bibitem{Schneider:2014mla}
  C.~Schneider and R.~Sch\"utzhold,
  JHEP {\bf 1602} (2016) 164
  [arXiv:1407.3584 [hep-th]].

\bibitem{Dumlu:2015paa}
  C.~K.~Dumlu,
  Phys.\ Rev.\ D {\bf 93} (2016) no.6,  065045
  [arXiv:1507.07005 [hep-th]].

\bibitem{Ilderton:2014mla}
  A.~Ilderton,
  JHEP {\bf 1409} (2014) 166
  [arXiv:1406.1513 [hep-th]].

\bibitem{Monin:2010qj}
  A.~Monin and M.~B.~Voloshin,
  Phys.\ Rev.\ D {\bf 81} (2010) 085014
  [arXiv:1001.3354 [hep-th]].

\bibitem{Satunin:2013an}
  P.~Satunin,
  Phys.\ Rev.\ D {\bf 87} (2013) no.10,  105015
  [arXiv:1301.5707 [hep-th]].

\bibitem{Schubert:2000yt}
  C.~Schubert,
  Nucl.\ Phys.\ B {\bf 585} (2000) 407
  [hep-ph/0001288].

\bibitem{Schubert:2001he}
  C.~Schubert,
  Phys.\ Rept.\  {\bf 355} (2001) 73
  [hep-th/0101036].

\bibitem{Gies:2011he}
  H.~Gies and L.~Roessler,
  Phys.\ Rev.\ D {\bf 84} (2011) 065035
  [arXiv:1107.0286 [hep-ph]].

\bibitem{Ilderton:2016qpj}
  A.~Ilderton and G.~Torgrimsson,
  Phys.\ Rev.\ D {\bf 93} (2016) no.8,  085006
  [arXiv:1601.05021 [hep-th]].

\bibitem{Dinu:2013gaa}
  V.~Dinu, T.~Heinzl, A.~Ilderton, M.~Marklund and G.~Torgrimsson,
  Phys.\ Rev.\ D {\bf 89} (2014) no.12,  125003
  [arXiv:1312.6419 [hep-ph]].

\bibitem{Meuren:2014uia}
  S.~Meuren, K.~Z.~Hatsagortsyan, C.~H.~Keitel and A.~Di Piazza,
  Phys.\ Rev.\ D {\bf 91} (2015) no.1,  013009
  [arXiv:1406.7235 [hep-ph]].

\bibitem{Bamber:1999zt}
  C.~Bamber {\it et al.},
  Phys.\ Rev.\ D {\bf 60} (1999) 092004.



\end{thebibliography}
\end{document}